# MORE ON THE INTERACTION REGION OF PEP-N*

M. Sullivan**, SLAC, Stanford, CA 94025, USA

## Abstract

The PEP-N project [1,2] consists of a small, very low-energy e− storage ring (VLER) located in one of the interaction-straight regions of PEP-II. The small ring is brought into collision with the low-energy (3.1 GeV) e+ beam (LER). The center-of-mass energies from this collision are between the $\Phi$ and $J/\psi$ resonances. We achieve a head-on collision through the use of a central magnetic dipole field that generates a large horizontal bending field. This field is also the central field of the detector. The large energy range of the VLER, in order to maximize the center-of-mass energy range, complicates the collision point geometry. In order to maintain the beam orbits near the collision point two techniques are used. The first is to scale the central dipole field up and down with the energy of the VLER and the second is to use passive shielding to decrease the integral B·dl of the dipole field seen by the VLER. Changes in the orbit of the LER are corrected with local bending magnets. Further details of the interaction region geometry as well as design issues that include synchrotron radiation from the high-current positron beam are discussed.

## 1 INTRODUCTION

The SLAC, LBNL, LLNL, PEP-II B-factory[3] consists of two storage rings located one above the other in the PEP tunnel at SLAC. The low-energy ring (LER) of positrons is 89 cm above the high-energy ring (HER) of electrons. At design operation, the LER beam current is 2.14 A in 1658 bunches that are 1.26 m apart. The HER beam current is 0.75 mA. The PEP-N proposal consists of a small storage ring of electrons that collide with the LER to produce a center-of-mass energy ($E_{cm}$) between the $\Phi$ and the $J/\psi$ (approximately 1 GeV to 3 GeV). The very low-energy (VLER) electron storage ring would be located in one of the old interaction region halls of PEP-I (interaction region 12 is presently being considered).

## 2 INTERACTION REGION DESIGN

Three major collision designs were considered: 1) head-on, 2) a small angle collision, similar to KEKB, 3) and a very large angle collision (> 100 mrads). The very large angle collision design would be an interesting accelerator to build but it was considered very risky and had a high probability of not producing the required luminosity as well as introduce perturbations on the LER that might adversely affect B-factory running. The smaller angle collision (~20 mrad total angle) presents two difficulties. One is that the magnetic elements needed for the VLER would take up most of the small angle acceptance of the detector. This is important since the energy asymmetry in PEP-N is very large over most of the $E_{cm}$ range of interest. The LER energy is held constant at 3.1 GeV while the VLER has an energy range of about 100-800 MeV. The detector angular acceptance in the forward boost direction is 100 mrad along the beam direction. In addition, a crossing angle collision means that the beam must be brought back over the LER beam in order to keep the small storage ring on one side of the LER beam line. This is especially difficult because it has to be done very soon after the collision. There is not much space in the ±10m long interaction region hall to get the beam back to the other side of the LER. This left the head-on solution as the best choice for the collision. The beams are brought into collision and separated by a large horizontal dipole field located at the interaction point that insures that the VLER stays on the same side of the LER beam line. This same magnetic field serves as the detector field.

The present working design uses a field model that has a maximum field strength of 3 kG. Figure 1 is a plot of the field from this magnet along the z axis.

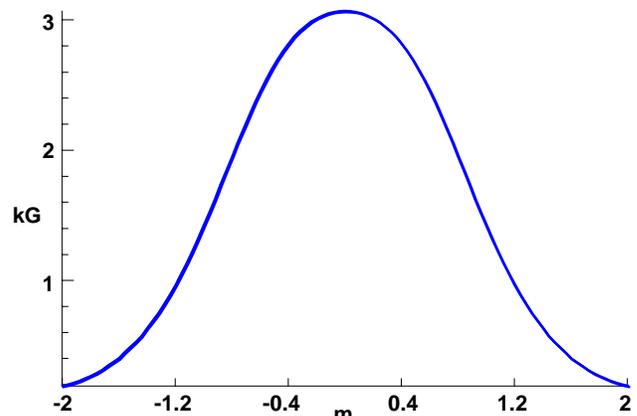

Figure 1. Plot of the magnetic field from the central dipole magnet. The field extends out to at least 2 m from the center. We will shield the accelerator beam pipes as much as possible to minimize the integrated strength of this magnet.

The IP is located −25 cm from the center of the field. The 3 kG field corresponds to an accelerator design that has a 557 MeV beam energy for the VLER. Placing the collision point −25 cm from the center of the magnet increases the amount of detector field in the boost direction and minimizes the amount of upstream magnetic field. This lowers the amount of upstream bending (and hence synchrotron radiation) in the LER.

## 2.1 Downstream beam lines

The downstream side of the collision point is in the direction of the LER. This is the side where most of the physics particles go. On this side, the VLER is deflected horizontally 192 mrads while the LER is deflected 34 mrads. This results in a separation of 26.7 mm between the two beams at the first parasitic crossing, 0.63 m from the IP, which translates into 38 $\sigma_x$ for the VLER and 68 $\sigma_x$ for the LER. This large separation makes any beam-beam effect from the parasitic crossing negligibly small. The beams are separated enough to allow each beam to enter a separate beam pipe about 1.3 m from the IP.

The first accelerator element after the central dipole field is a vertically focusing quadrupole (QDI1) for the VLER located 1.5-1.7 m from the IP. QDI1 is constructed from permanent magnet material. The compactness of the permanent magnet design permits this magnet to be 1.5 m from the IP and yet not have any effect on the nearby LER beam. The small design also maximizes the solid angle acceptance of the detector. Following QDI1 is a horizontal bending magnet (B0VL). This magnet starts the reverse bend on the VLER that brings the VLER back parallel to the LER. The next VLER element is a horizontally focusing quadrupole (QFI2) located 2.5-2.8 m from the IP. This magnet is far enough away to no longer interfere with the LER and it has only a minor impact on the detector solid angle. Following QFI2 is another quad QDI3 from 3.3-3.6 m. A reverse bend horizontal dipole (B1VL) at 3.7-4.1 m straightens out the VLER orbit to be parallel to and 40 cm from the LER followed by one more matching quad (QFI4) at 4.2-4.5 m.

The downstream LER beam line includes 3 horizontal dipole bend magnets to correct the orbit back to the nominal trajectory and to match dispersion.

## 2.2 Upstream beam lines

The upstream side of the IP has very similar magnet placement as the downstream side, however the beams are not separated as quickly so the implementation is different. On this side the beam separation at the first parasitic crossing is 22.8 mm which is $32\sigma_x$ for the VLER and $58\sigma_x$ for the LER, still large enough to make tune shifts from this parasitic crossing negligible. Moving out from the IP we find that the first accelerator element is a horizontal dipole magnet (1-1.3

m) that both beams travel through. This magnet can be used to add or subtract to the central dipole field and is used to maintain the VLER orbit when the VLER energy is changed. The next element is QDI1 (1.5-1.8 m). However, on this side of the IP this large aperture magnet is seen by both beams and is a normal steel magnet. The center of this horizontally defocusing magnet is positioned close to the LER beam minimizing the bend for this beam and maximizing the bend for the VLER. The extra horizontal kick from this magnet separates the beams enough so that the next element (QFI2, 2.5-2.8 m) can be a septum quadrupole with a field-free drift region for the LER. The rest of the magnets in the VLER are essentially the same as the downstream side with the VLER beam parallel to and offset from the LER design trajectory by 40 cm 4.1 m from the IP. The LER beam line includes four horizontal dipole magnets to steer the LER back to the nominal orbit and to close dispersion. Figures 2 and 3 are layout pictures of the interaction region.

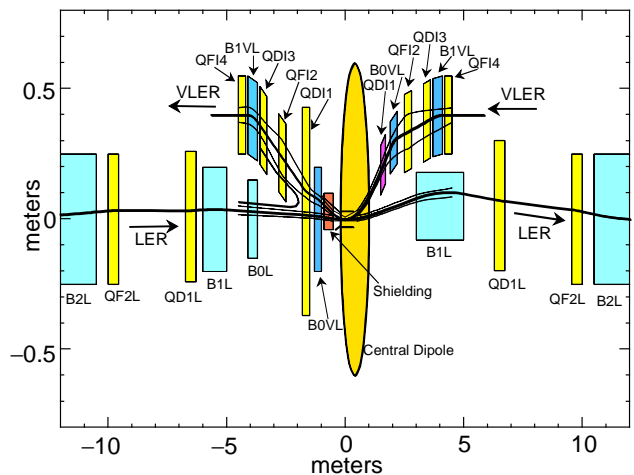

Figure 2. Layout of the interaction region. Please note the exaggerated left-hand scale.

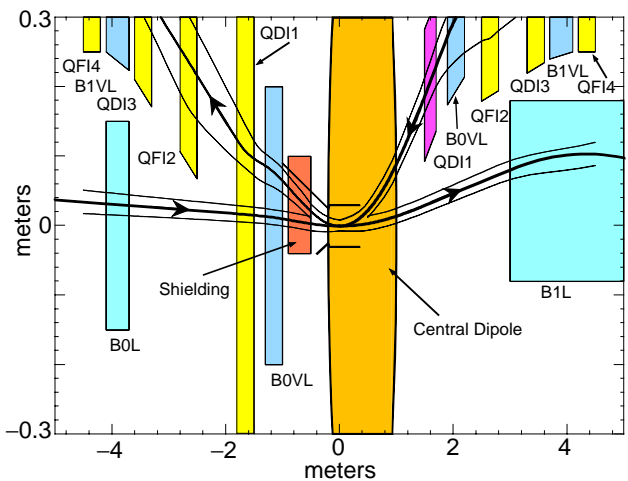

Figure 3. Close up of the interaction region.

# 3 CHANGING THE CENTER-OF-MASS ENERGY OF PEP-N

The LER energy is fixed by PEP-II at 3.1 GeV. In order to change the Ecm we must change the energy of the VLER. The interaction region baseline design of PEP-N has a 557 MeV VLER and a 3 kG central dipole field. In order to reach the J/ψ resonance the VLER energy must be increased to 780 MeV. In order to maintain the beam orbits and get the beams to separate properly we proportionally increase the field of the central dipole to 4.2 kG. Decreasing the VLER energy from the 557 MeV deign point is done differently. The detector collaboration prefers a higher central magnetic field while the accelerator designers prefer a lower central field. With this in mind the present design tries to maintain 3 kG as a minimum value for the central field. Therefore, in order to lower the VLER energy and maintain the central field at 3 kG, passive shielding is added to the beam pipe to subtract some of the central field from the VLER beam. With this technique, we achieve a VLER energy of 347 MeV.

In order to go still lower in VLER energy, the present strategy would be to rebuild the beam pipes in the interaction region allowing for a much larger angle of separation at 347 MeV through the use of the unshielded central field of 3 kG. Once again, we would add passive shielding to the VLER beam pipe as the VLER energy is lowered from the 347 MeV starting point. We note here that this strategy is still preliminary and further refinements will no doubt be forth coming.

# 4 SYNCHROTRON RADIATION

The 780 MeV VLER design has the highest levels of synchrotron radiation. However, the fact that the IP is

−25 cm upstream from the center of the main field and that we use the offset QDI1 magnet to further separate the beams means that the upstream LER has relatively weak bending magnets. The main source of synchrotron radiation power comes from the two closest of the four dipole magnets on the LER beam line. The strength of these magnets for the 780 MeV VLER is 2.1 and 2.0 kG and they generate 465 and 1054 W respectively with a 2.14 A LER beam. The critical energies of these bend magnets are 1.36 and 1.3 keV. The power levels are low enough to not pose a problem for beam pipe cooling. More complete studies need to be made, but synchrotron radiation power does not seem to be an issue and the very low critical energies argue that detector background levels from synchrotron radiation will be low. Figure 4 shows the fan of radiation coming from these magnets.

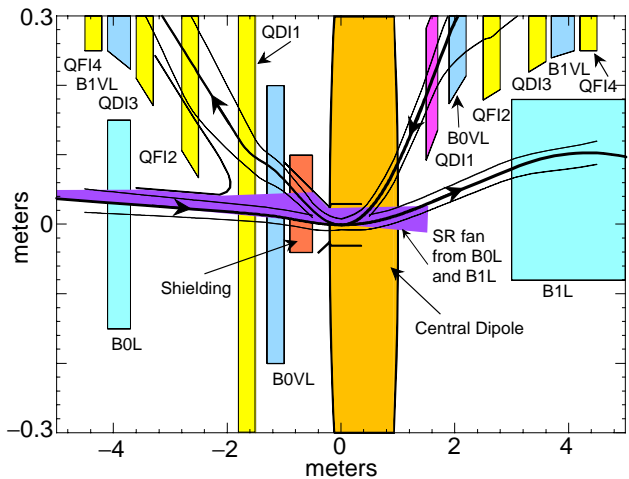

Figure 4. Interaction region layout with the synchrotron radiation fan from the B0L and B1L magnets of the LER. The fan strikes some of the upstream beam pipes. The low critical energies of this bending radiation argues that synchrotron radiation backgrounds will not be an issue but a more complete study needs to be made to insure that backgrounds are acceptable.

# 5 SUMMARY

The interaction region design of PEP-N employs a central horizontal dipole field to bring the beams into collision and serve as the central detector field. The collision point is shifted −25 cm upstream in the LER in order to maximize the detector field in the forward direction where most of the physics particles travel as well as minimize the upstream bending magnet strengths that contribute to detector background issues. A baseline design with a VLER energy of about 500 MeV has a 3 kG central field. In order to move the $E_{cm}$ up to the J/ψ we increase the VLER energy up to 780 MeV and scale up the central field to 4.2 kG. This maintains the orbit separation geometry. Lowering the VLER energy below the 557 MeV baseline energy is accomplished by adding passive shielding around the beam pipe. This allows the central field to remain at 3 kG down to a VLER energy of 347 MeV. Lowering the VLER energy further involves modifying the orbit separation geometry.

The central dipole field separates the beams so efficiently ($>50\sigma_x$) that there are virtually no beam-beam effects from parasitic crossings ±63 cm from the IP. In fact, there is even enough separation ($>14\sigma_x$) at the halfway point (31.5 cm) to allow the VLER bunches to be positioned there which means that the beams can be moved rapidly out of collision by a change in the VLER RF phase.